\documentclass[conference]{IEEEtran}

\usepackage{cite}
\usepackage{color}
\usepackage{url}

\ifCLASSINFOpdf
   \usepackage[pdftex]{graphicx}
\else
   \usepackage[dvips]{graphicx}
   \graphicspath{{../eps/}}
   \DeclareGraphicsExtensions{.eps}
\fi

\begin{document}
%
\title{Automatic link extraction:\\
The good, the bad and the ugly in software ecosystem mining}


\author{\IEEEauthorblockN{Eleni Constantinou and Tom Mens}
\IEEEauthorblockA{Software Engineering Lab, COMPLEXYS Research Institute\\
University of Mons, Belgium\\
first.last@umons.ac.be}
}

\maketitle

%
\IEEEpeerreviewmaketitle

\section{Introduction}
\label{sec:introduction}
Software ecosystems are coherent collections of evolving interdependent software projects that are developed and maintained by a community of interacting contributors. Each project within an ecosystem uses different online collaboration platforms throughout its lifetime, either to support different aspects of the development process (e.g., version control, bug tracking, documentation, communication, user feedback) or due to migration from one platform to a competitive one.

To gain a complete view of an ecosystem's software history, the evolutionary data of all its constituent projects must be retrieved from these different platforms. A necessary pre-processing step is to extract, for each project belonging to the ecosystem, the links to its repositories stored in the different collaborative platforms~\cite{Constantinou:2017:ISSE,Decan2016,Kikas:2017}. For example, software package managers such as npm and RubyGems provide metadata containing links provided by the developers pointing to the source code repository, mailing list, bug tracker and so on. Given the fact that software ecosystems consist of thousands of projects, researchers usually focus on large-scale analyses and this pre-processing step is automated.

While such automated link extraction can retrieve valid links, it often also includes false positives, such as links that do not correspond to a valid repository. It may also result in false negatives, e.g., links that could have helped in recovering a valid project repository, but for which the project repository is not explicitly stated. Thus, the process of automatic link extraction is prone to errors or omissions.
Based on our previous research~\cite{Constantinou:2017:ISSE,Constantinou:2017} we observed that the majority of projects (close to 55\%) within the npm and RubyGems software ecosystems provide valid links. However, omitting relevant repositories or including non-relevant repositories can mislead subsequent analyses.

This work presents the automatic link extraction pitfalls based on our experience on manually investigating links in the RubyGems package manager metadata. This work can lead in automating the link extraction approach so as to avoid these pitfalls and produce more complete datasets to be used by researchers when they investigate the multi-platform evolution of software ecosystems.

\section{Data Fields}
\label{sec:data_fields}
When parsing the metadata provided by the RubyGems API, there are several available fields 
where links can be found containing project-specific data: project\_uri, homepage\_uri, wiki\_uri, documentation\_uri, mailing\_list\_uri, source\_code\_uri and bug\_tracker\_uri.

Based on our experience with RubyGems data extraction, we found that one cannot assume such data to be complete in practice, e.g., gems only specify a bug tracker link, which can be a link to the issues mechanism of their GitHub source code repository.
Therefore, we do not target specific fields, rather parse all the fields to recover links to either GitHub or other domains. Also, we examine links to domains other than GitHub in the investigation of each gem, since they might lead to a company's website which maintains those gems or a developer's personal webpage where a link to his/her GitHub account can be found.

\section{The good}
\label{sec:the_good}
The best case scenario happens when a gem specifies an explicit and correctly structured link to a GitHub repository. Regardless of the field in which the gem maintainers set the GitHub repository, in these cases they provide an explicit and valid link to their source code repository.

\section{The bad}
\label{sec:the_bad}
In many cases, however, the links provided through the uri fields can be considered as ``bad'' in the sense that they do not point to an expected data source. We classified such bad links under the following sub-categories:

\textbf{Irrelevant links:} During our analysis of RubyGems, we found a wide variety of irrelevant links, including pointers to Chinese websites similar to SlideShare, web pages, and irrelevant or overly generic websites (e.g., \url{www.google.com}, \url{http://php.net/}) or GitHub (\url{www.github.com}). In these cases no further investigation of the links takes place.

\textbf{Invalid GitHub links:} Seemingly valid links that do not point to a valid GitHub repository.  This may happen because the project's GitHub repository has been renamed. In that case we store the link to the renamed repository (obtained by parsing the project's information with the GitHub API). It may also happen that the project has been removed from GitHub, in which case we discard the link.

\textbf{GitHub pages:} GitHub links that lead to the webpage of the developer rather than pointing to the project itself, e.g., \textsf{username.github.io}. For those cases, the link of the GitHub page can be transformed to formulate a candidate repository link (e.g., \url{https://github.com/username/gem_name}). After manually validating the repository as the official gem repository, we can include it in our dataset.

\textbf{Sub-directory GitHub links:} Valid links that point to a sub-directory of the project repository.  We ignore these cases since such repositories can either include the development history of multiple gems or the development of a single gem alongside other components that are irrelevant to RubyGems. 

\textbf{GitHub user profile} (e.g., \url{https://github.com/username}): If we encounter such a link, we formulate a candidate GitHub link in the form \url{https://github.com/username/gem_name} and validate if this repository hosts the development of the gem.

\section{The ugly - Implicit links}
\label{sec:the_ugly}
``Ugly" links concern non-GitHub links that point to personal websites or company webpages. These cases are taken into account since these links may point to information that can help in retrieving the project's GitHub repository. For each such link, we visit the external webpage (often a personal or company webpage) and try to recover a link to a corresponding GitHub user or organization. If such a link is found, then we search the repositories of this user/organization to find a repository with the same name of the gem. If one is found, we validate that it hosts the gem's source code repository.

If we cannot find a link to a GitHub user or organization, we try to infer one by checking if the link \url{https://github.com/username} exists: If it doesn't exist, our search strategy stops; otherwise we confirm that this account corresponds to the developer or company by searching for links pointing back to the homepage and in the case of companies, by also comparing if the logo of the company matches. If these criteria are met, then we resume our repository search strategy.

To illustrate this process, consider the following example. By visiting the homepage uri\footnote{\url{http://www.futureworkshops.com}} of the \emph{notifiable-rails} gem, we couldn't find a link a GitHub account or repository, so we tried finding a GitHub organization with the same name \url{https://github.com/futureworkshops}. We confirmed that this is indeed the right GitHub organization since it provides a link back to the company's webpage and it contains the same logo. Next, by searching for a repository with the same name as the gem, we found the \url{https://github.com/FutureWorkshops/notifiable-rails} link and by checking the gem specification file, we confirmed that it is indeed the notifiable-rails repository.

\section{Discussion}
This work focuses on extracting GitHub links. However, some projects provide non-GitHub links which correspond to other software repositories (SourceForge, GitLab, BitBucket). The aforementioned search strategies also apply on the search strategy of these platforms since the pitfalls of extracting the links are similar to the ones when extracting GitHub links. Therefore, explicit links to such platforms would fall under the ``Good" category, unless implicit or malformed links are provided.

Based on our experience with RubyGems, we found that the most appropriate regular expression to extract valid GitHub links (excluding multi-gem or multi-component repositories) is:

\vspace{0.1cm}
\fbox{\begin{minipage}{0.95\columnwidth}
\^{}https?(://github.com)((/)[\^{}/]+)\{2,2\}(.git$|/$)?\$\\
Issues:\\
\^{}https?(://github.com)((/)[\^{}/]+)\{2,2\}(/issues)($/$)?\$
\end{minipage}}
\vspace{0.1cm}

Information about deleted or renamed repositories can be retrieved through the GitHub API by requesting information about the repository; if it is renamed the new links are retrieved and if it is deleted a not found response is retrieved. Relying on available datasets to find the GitHub links (e.g., GHTorrent~\cite{Gousios13}) can impact the link extraction outcome since repositories might have been deleted while marked as valid in the dataset or the link to the renamed repository might not be linked to the old repository name. Also, based on our results, we found that GHTorrent does not contain valid repositories that are still active on GitHub.

\section{Conclusion}
\label{sec:conclusion}
Overall, our past experience with link extraction can assist in formulating a general link extraction process to match repositories across different platforms. Once formulated, this process can be validated based on our existing manual analysis results of RubyGems, as well as the ongoing link extraction in the npm ecosystem.

\section*{Acknowledgment}
This research was carried out in the context of FNRS cr\'edit de recherche J.0023.16 entitled ``Analysis of Software Project Survival'' and the bilateral collaborative research program FRQ-FNRS 30440672 entitled ``Towards an Interdisciplinary Socio-Technical Methodology and Analysis of Software Ecosystem Health''.

\bibliographystyle{IEEEtran}
\bibliography{ecbib}

\end{document}